\documentclass[aps,prd,12point,nofootinbib,showpacs,superscriptaddress]{revtex4-1}
\def\theequation{\arabic{section}.\arabic{equation}}
\usepackage{psfrag}
\usepackage{subfigure}
\usepackage{color}
\usepackage{mathrsfs}
\usepackage{graphicx}
\usepackage{amssymb, bm}
\usepackage{amsmath, amsthm}
\usepackage{epstopdf}
\usepackage{hyperref}
\usepackage{enumerate}
\usepackage{longtable}
\usepackage{amsmath}
\usepackage{amsfonts} % needed for bold Greek, Fraktur, and blackboard bold
\usepackage{graphicx} % needed for figures
\usepackage[normalem]{ulem}
\usepackage{amsfonts}
\usepackage[T1]{fontenc}
\usepackage{soul}

\newcommand{\be}{\begin{equation}}
\newcommand{\ee}{\end{equation}}
\begin{document}
\def\theequation{\arabic{section}.\arabic{equation}} 
%\preprint{}

\title{Analogies between logistic equation and relativistic cosmology} 



\author{Steve Dussault}
%\email[sdussault19@ubishops.ca]{}
%\thanks{}
\affiliation{Department of Physics \& Astronomy, Bishop's University, 2600 
College Street, Sherbrooke, Qu\'ebec, Canada J1M~1Z7
}

\author{Valerio Faraoni}
%\email[vfaraoni@ubishops.ca]{}
%\thanks{}
\affiliation{Department of Physics \& Astronomy, Bishop's University, 2600 
College Street, Sherbrooke, Qu\'ebec, Canada J1M~1Z7
}

\author{Andrea Giusti}
%\email[vfaraoni@ubishops.ca]{}
%\thanks{}
\affiliation{Institute for Theoretical Physics, ETH 
Zurich,cWolfgang-Pauli-Strasse 27, 8093, Zurich, Switzerland}


%\date{\today}

\bigskip
\bigskip
\begin{abstract}

We develop several formal analogies between the logistic equation and the 
spatially homogeneous and isotropic relativistic cosmology described by 
the Einstein-Friedmann equations. These analogies produce an effective 
Lagrangian and Hamiltonian and new symmetries for the logistic equation.

\end{abstract}

%\pacs{}
%\keywords{}

\maketitle

\section{Introduction}
\setcounter{equation}{0}
\label{sec:1}

The logistic equation is
\be
\dot{f}(t)=rf(t) \left[ 1-f(t) \right] \,,\label{logistic}
\ee
where $r$ is a positive constant and an overdot denotes differentiation 
with respect to $t$. The solution is the well known sigmoid
\be
f(t)= \frac{f_0 \, \mbox{e}^{rt}}{1+f_0 \, \mbox{e}^{rt}} 
\,,\label{sigmoid}
\ee
where $f_0$ is a constant. 
If $f_0 <-1$, the initial value at $t=0$ is $f(0)= \frac{f_0}{1+f_0}= 
\frac{|f_0|}{|f_0|-1}>1$ and $\dot{f}<0$. This solution decreases and 
approaches $1$ asymptotically. If instead $f_0>0$, then $f(0) <1$ and the 
solution always satisfies $f<1$. We require $f\geq 0$ and we restrict to 
the range $ 0< f \leq 1$.

Other solutions are the constants $f\equiv 1$, which is a fixed point 
solution and a late-time attractor, and $f(t) \equiv 0 $ which is also is 
also a fixed point and a repellor but will not be considered since we 
restrict to positive $f$.

The logistic equation and its applications are well known: here we point 
out a novel aspect of it, {\em i.e.}, formal analogies between the 
logistic equation and the Friedmann equation of spatially homogenous and 
isotropic cosmology in general relativity. There are two analogues, 
corresponding to the two different time coordinates widely used in 
cosmology. First, these analogies generate an effective Lagrangian and 
Hamiltonian for the logistic equation or, in other words, solve its 
inverse Lagrangian problem. Second, the Friedmann equation admits certain 
symmetries which can be used to generate corresponding symmetries on the 
logistic equation side of the analogy.

In the next section we recall the basics of spatially homogeneous and 
isotropic cosmology (also called Friedmann-Lema\^itre-Robertson-Walker or, 
in short, FLRW, cosmology) in Einstein's theory of gravity and the 
symmetries of the corresponding Einstein-Friedmann equations. In the 
following section, we develop the formal analogies between logistic 
equation and Friedmann equation, written using comoving time or conformal 
time, respectively. In both cases, we write down the corresponding 
Lagrangian and Hamiltonian for the logistic equation, and examine whether 
the symmetries of the Einstein-Friedmann equations generate new symmetries 
of the logistic equation. The last section contains the conclusions.

\section{Basics of FLRW cosmology}
\setcounter{equation}{0}
\label{sec:2}

The foundation of relativistic cosmology is the Copernican principle, 
according to which the universe is spatially homogeneous and isotropic on 
large scales ({\em i.e.}, averaging over spatial regions of size larger 
than approximately 100 Megaparsecs) 
\cite{HawkingEllis,Wald,Carroll,KolbTurner,LiddleLyth,Liddle}.  
Therefore, the relativistic spacetime manifold used to model the universe 
is taken to satisfy the symmetry requirements of spatial homogeneity and 
isotropy. The latter are quite stringent: in fact, the spacetime manifold 
must have constant spatial curvature \cite{Eisenhart}.  The spacetime 
geometry must then necessarily be the one described by the FLRW metric 
tensor. Using polar comoving coordinates\footnote{The comoving coordinate 
are associated with observers comoving with the cosmic fluid, {\em i.e.}, 
those that see the cosmic microwave background radiation homogeneous and 
isotropic around them (apart from tiny temperature perturbations of the 
order $ \delta T/T \sim 5 \times 10^{-5}$).}  $\left(t,r,\vartheta, 
\varphi \right)$, the FLRW line element is \cite{HawkingEllis,Wald}
\begin{equation}
ds^2 = -dt^2 + a^2(t) \Bigg( \frac{dr^2}{1-Kr^2} + r^2 d\Omega_{(2)}^2 
\Bigg)    \label{FLRWlineelement}
\end{equation}
where $d\Omega_{(2)}^2 = d\vartheta^2 + \sin^2 \vartheta \, d\varphi^2$ 
denotes  the line 
element on the unit 2-sphere, while the sign of the constant ``curvature 
index''  $K$ characterizes  the 3-dimensional spatial sections 
corresponding to instants of comoving time $t=$~const. A positive 
curvature index $K>0$ is associated with 3-spheres; vanishing curvature 
$K=0$ corresponds to Euclidean spatial sections, while an index $K<0$ is 
associated with  hyperbolic 3-dimensional spatial sections.

\subsection{Einstein-Friedmann equations}

Thanks to the symmetries, the entire evolution of FLRW 
universes is described by the functional form of the scale factor $a(t)$. 
The spacetime geometry is curved by some matter source: in FLRW cosmology, 
it is customary (but not compulsory) to use as the  
matter source  a perfect fluid with energy density $\rho(t)$ and 
pressure given by $P(t)=w\rho(t)$. Here $w$ denotes a constant equation of 
state 
parameter. The evolution of the cosmic scale factor $a(t)$ and energy 
density $ \rho(t)$  is ruled by the 
Einstein equations, which are greatly simplified by the high degree of 
symmetry of FLRW cosmology and reduce to a set of three ordinary 
differential equations, the Einstein-Friedmann equations. These equations 
include \cite{HawkingEllis,Wald,Carroll} the 
Friedmann equation
\begin{equation}
H^2 =   \frac{8 \pi G}{3} 
\rho - \frac{K}{a^2} +\frac{\Lambda}{3} \label{eq:Friedmann}
\end{equation}
(which amounts to a first order constraint on the dynamics);  the 
acceleration (or Raychaudhuri) equation 
\begin{equation}
\frac{\ddot{a}}{a}= -\frac{4 \pi G}{3} \left( \rho + 3 P \right) +  
\frac{\Lambda}{3} \,;\label{eq:acceleration}
\end{equation}
and the equation describing covariant conservation of energy of the cosmic 
fluid
\be
\dot{\rho}+3H \left( P+\rho \right)=0  \,.\label{eq:conservation}
\ee
The constants appearing in these equations are Newton's constant $G$ and 
Einstein's celebrated cosmological constant $\Lambda$. We use an overdot 
to indicate differentiation with respect to the comoving time $t$, while 
$H \equiv \dot{a}/a$ is the Hubble function \cite{Wald,Carroll}. The 
cosmological constant $\Lambda$ can be treated formally as an effective 
perfect fluid with energy density and pressure $ \rho_{\Lambda}=\frac{ 
\Lambda}{8\pi G} =-P_{\Lambda}$ 
\cite{Wald,Carroll}.  Following the notation of Refs.~\cite{Wald,Carroll}, 
we use units in which the speed of light is unity. 

For a generic cosmic fluid, only two out of the three 
equations~\eqref{eq:Friedmann}-\eqref{eq:conservation} are independent and 
one of them can be derived from the other two, if the Einstein equations 
and the expression of the Ricci scalar in FLRW space are used as extra 
information. Specifically:

\begin{itemize}

\item Differentiating the Friedmann equation~(\ref{eq:Friedmann}) with 
respect to time yields
\be
2H\left( \frac{\ddot{a}}{a} -H^2\right) =\frac{8\pi G}{3} \, \dot{\rho} 
+\frac{2KH}{a^2} \,;\label{somelabel1}
\ee
using the acceleration equation~(\ref{eq:acceleration}) and the Friedmann 
equation~(\ref{eq:Friedmann}) to substitute for $\ddot{a}/a$ and $H^2$ 
in~(\ref{somelabel1}) produces the conservation 
equation~(\ref{eq:conservation}). This derivation of 
Eq.~(\ref{eq:conservation}) is a reflection of the 
more general derivation of the covariant conservation equation $\nabla^b 
T_{ab}=0$ from the Einstein equations
\be
R_{ab}-\frac{1}{2} \, g_{ab} R +\Lambda g_{ab}=8\pi G T_{ab} 
\label{Einsteineqs}
\ee
and the contracted Bianchi identites $\nabla^b \left(R_{ab}- g_{ab} R 
/2\right)=0$ (here $g_{ab}$, $R_{ab}$, and $T_{ab}$ are the metric 
tensor, its Ricci tensor, and the stress-energy tensor of matter, 
respectively, and $R\equiv g^{ab}R_{ab}$).

\item The Friedmann equation~(\ref{eq:Friedmann}) follows from the 
acceleration equation~(\ref{eq:acceleration}), the Einstein 
equations~(\ref{Einsteineqs}), and the expression of the 
Ricci scalar in the FLRW geometry~(\ref{FLRWlineelement})
\be
R=6 \left( \dot{H}+2H^2 +\frac{K}{a^2} \right) \,.  \label{eq:R}
\ee
In fact, a perfect fluid stress-energy tensor
\be
T_{ab} =\left( P+\rho \right) u_a u_b +P g_{ab} 
\ee
(where $u^c$ is the fluid four-velocity normalized to $u^c u_c=-1$) has 
trace $T=-\rho+3P$ and the contraction of the Einstein 
equations~(\ref{Einsteineqs}) then gives
\be
R=8\pi G \left( \rho-3P\right) +4\Lambda \,. \label{contraction}
\ee
Equating the right hand sides of Eqs.~(\ref{eq:R}) and 
~(\ref{contraction}) yields
\be
\frac{\ddot{a}}{a} +H^2 +\frac{K}{a^2} =\frac{4\pi G}{3} \left( 
\rho-3P\right) +\frac{2\Lambda}{3} \,.
\ee 
Using now the acceleration equation~(\ref{eq:acceleration}) to substitute 
for $\ddot{a}/a$, one obtains the Friedmann equation~(\ref{eq:Friedmann}).

\item The acceleration equation~(\ref{eq:acceleration}) can be derived 
from the Friedmann equation~(\ref{eq:Friedmann}) and the 
energy conservation equation~(\ref{eq:conservation}). In fact, 
differentiating~(\ref{eq:Friedmann}) with respect to time leads to 
\be
2H\left( \frac{\ddot{a}}{a} -H^2 \right) = \frac{8\pi G}{3} \, \dot{\rho} 
+ \frac{2KH}{a^2} \,;
\ee
using Eq.~(\ref{eq:conservation}) to substitute for $\dot{\rho}$ and the 
Friedmann equation~(\ref{eq:Friedmann}) for $H^2$, the 
acceleration equation~(\ref{eq:acceleration}) follows straightforwardly.

\end{itemize}

By specifying a constant 
equation of state $P=w\rho$, the conservation equation is immediately 
integrated to
\be
\rho(a)= \frac{\rho_0}{ a^{3(w+1)} } \,.
\ee
Analytic solutions of the Einstein-Friedmann equations 
(\ref{eq:Friedmann})-(\ref{eq:conservation}) with a single perfect fluid, 
as well as analyses of their phase space, are well known (see 
Refs.~\cite{oldAmJP,myAmJP,SonegoTalamini} for reviews).

\subsection{FLRW Lagrangian and Hamiltonian}

A Lagrangian for FLRW cosmology is 
\be
L\left(a, \dot{a} \right) = 3a\dot{a}^2 + 8\pi G a^3 \rho -3Ka\,, 
\label{FLRWlagrangian}
\ee
where $\rho=\rho(a)$ is specified by the barotropic  equation of state 
$P=P(\rho)$ and the conservation equation 
$\dot{\rho}+3H(P+\rho) =0$. Since $\rho=\rho(a)$, the  
Lagrangian~(\ref{FLRWlagrangian}) does not depend explicitly on $t$ 
and the corresponding Hamiltonian is conserved,
\be
{\cal H}=\frac{\partial L}{\partial \dot{a}} \, \dot{a} -L=  
3a\dot{a}^2 -8\pi G a^3 \rho +3Ka = C \,. 
\ee
These Lagrangian and Hamiltonian solve the inverse variational problem of 
finding an action integral for the logistic equation, however this is not 
sufficient to complete the derivation.  Since the dynamics of general 
relativity is constrained  
\cite{Wald,Carroll}, in the FLRW geometry the Hamiltonian 
constraint (time-time component of the Einstein equations) imposes 
that $C=0$ and is the same as the Friedmann equation~(\ref{eq:Friedmann})  
\cite{Wald}.

\subsection{Symmetries of the Einstein-Friedmann equations for spatially 
flat universes}

If the universe is spatially flat ($K=0$) and there is a single term 
in the right hand side of the Friedmann equation~(\ref{eq:Friedmann}), 
corresponding to a single perfect fluid with constant equation of state 
$P=w\rho$,  the 
Einstein-Friedmann equations exhibit certain symmetries \cite{Chimento, 
myPLB,Symmetry2019} which are studied in the cosmological literature for 
the purpose of generating analytic solutions 
\cite{Marek1, Marek2,ParsonsBarrow,Chimento, myPLB,BarrowPaliathanasis, 
Symmetry2019,Pailasetal20, Pailas2, Pailas3, Pailas4, Pailas5, Pailas6, 
Pailas7, Pailas8, Pailas9, Pailas10, Pailas11, Hobson, Achour1, Achour2, 
Achour3}. 
In these symmetry 
operations, time $t$, scale 
factor $a$, or Hubble function $H$ are rescaled and the cosmic fluid is 
changed appropriately, leaving the Einstein-Friedmann equations 
invariant in form. 

Under the first symmetry \cite{myPLB} 
\begin{eqnarray}
a & \rightarrow & \tilde{a} = \frac{1}{a} \,, \label{symm1a}\\
&&\nonumber\\
w & \rightarrow & \tilde{w} = -(w+2) \,, \label{symm1b}
\end{eqnarray}
an expanding universe changes into a contracting one and {\em 
vice-versa}.\footnote{When  it is applicable to the logistic equation, 
this symmetry will change an increasing solution with $ f<1$ 
into a decreasing one with $f>1$.} 

Another one-parameter group of symmetry transformations of a $K=0$ 
universe is \cite{Symmetry2019}
\begin{eqnarray}
a &\rightarrow & \bar{a}=a^s \,,\label{symm2a}\\
&&\nonumber\\
dt & \rightarrow & d\bar{t} = s \, a^{ \frac{3(w+1)(s-1)}{2} } dt 
\,,\label{symm2b}\\
&&\nonumber\\
\rho & \rightarrow & \bar{\rho} = a^{ -3(w+1)(s-1)} \rho \,,\label{symm2c}
\end{eqnarray}
where the real number $s\neq 0$ parametrizes the transformation. These 
operations form a one-parameter Abelian group.

Finally, the third type of symmetry for a spatially flat FLRW cosmology is 
\cite{Chimento,DussaultFaraoni} 
\begin{eqnarray}
\rho & \rightarrow &  \bar{\rho}=\bar{\rho} (\rho) \,,\label{symm3a}\\
&&\nonumber\\ 
H & \rightarrow &  \bar{H}=\sqrt{ \frac{ \bar{\rho}}{\rho}  }  \, \, H 
\,,\label{symm3b}\\
&& \nonumber\\
P & \rightarrow &  \bar{P }=-\bar{\rho} + \sqrt{ \frac{ \rho}{\bar{\rho}} 
}  \, 
\left( P+\rho\right) \, \frac{ d\bar{\rho} }{d\rho} \,, \label{symm3c}  
\end{eqnarray}
where the function $\bar{\rho}(\rho)$ has the same sign as $\rho$ and is  
regular. This symmetry applies also when the barotropic perfect fluid in 
the FLRW universe does not have linear or constant equation of state.

\section{Cosmological analogies}
\setcounter{equation}{0}
\label{sec:3}

Dividing the logistic equation~(\ref{logistic}) by $f$ and squaring gives 
\be
\left( \frac{\dot{f}}{f} \right)^2 = r^2 -2r^2 f +r^2 f^2  
\,.\label{logisticreduced}
\ee
In the following we examine analogies with the Friedmann 
equation~(\ref{eq:Friedmann}).

\subsection{First analogy using comoving time}

Equation~(\ref{logisticreduced})  is formally analogous\footnote{In  FLRW 
cosmology, the scale factor can be rescaled by a  constant  
without affecting the physics. However, the solution $f(t)$ of  the 
logistic equation does not enjoy this property, because  of 
the quantity $1-f$ appearing in brackets in the right hand side of 
Eq.~(\ref{logistic}). By imposing 
the analogy between $a(t)$ and $f(t)$, one loses this property of the 
scale factor $a(t)$.} to the Friedmann equation~(\ref{eq:Friedmann}) 
with $\left( t, a(t) \right) \leftrightarrow \left( t, f(t) \right) $, 
$K=0$, $\Lambda =3r^2$, and energy 
density 
\be
\rho=\rho_0 a\left(a-2\right) \label{rho}
\ee
with $\rho_0=3r^2 /( 8\pi G)$. If the physical requirement that the 
energy density  be 
non-negative is imposed, the scale factor must satisfy the lower 
bound $a(t) \geq 2 $~ ~$\forall t$. 
However, one can in principle relax this assumption if one is only 
interested in  the mathematical properties of this formal 
analogy.  In order to complete 
the analogy, one must impose that another of the Einstein-Friedmann 
equations be satisfied. For simplicity, we choose to impose the covariant 
conservation 
equation $\dot{\rho}+3H\left( P+\rho \right)=0$, which then yields the 
effective equation of state of the cosmic fluid
\be 
P=-\rho -\frac{a}{3} \, \frac{d\rho}{da} \,. \label{prepressure}
\ee
Eq.~(\ref{rho}) can be sen as the quadratic $a^2 -2a-\rho/\rho_0=0$; 
for physical reasons  only the positive root 
\be
a=1+\sqrt{ 1+\frac{\rho}{\rho_0}}  \label{positiveroot} 
\ee
is relevant in cosmology, hence we choose the positive sign and we 
restrict ourselves to $f>0$ in the    
logistic equation. Substituting 
Eq.~(\ref{positiveroot}) into Eq.~(\ref{prepressure}) yields 
the nonlinear equation of state of the analogous cosmic fluid
\be
P=-\frac{5}{3} \, \rho -\frac{2\rho_0}{3}\left[ 1+  \sqrt{ 1+ 
\frac{\rho}{\rho_0}} \, \right]  \,,
\ee
which describes  a phantom fluid ({\it i.e.}, $P/\rho <-1$). To summarize, 
this analogous universe is spatially flat, has positive cosmological 
constant, and is filled with an exotic phantom perfect fluid. From a 
physical point of view, the 
energy density must be non-negative, which implies $a\geq 2$, while 
the sigmoid~(\ref{sigmoid}) is never 
larger than one. Therefore, $\rho < 0$ and one should discard this first 
analogy from the physical point of view. However, 
if only the mathematical properties are of interest, the formal analogy  
still stands.\\\\

\noindent {\em Lagrangian and Hamiltonian.} One can consider the inverse 
variational problem for Eq.~(\ref{logistic}). The cosmological analogy 
gives the Lagrangian for the logistic equation
\be
L_1\left( f, \dot{f} \right) = f\dot{f}^2 + r^2 f^4 \left( f-2 \right) 
+ r^2 f^3 \,. \label{L1}
\ee
Since $L_1$ does not depend explicitly on time, the corresponding 
Hamiltonian is conserved, 
\be
{\cal H}_1 = \frac{\partial L_1}{\partial \dot{f} }\, \dot{f} - L_1 
= f\dot{f}^2 -r^2 f^4 \left(f-2\right) -r^2 f^3 = C_1  \,. \label{H1}
\ee 
Setting the constant $C_1=0$ yields $ \dot{f} =\pm |r| f \left(1-f 
\right)$. 
The negative sign corresponds to inverting the sign of the constant $r$, 
which was stipulated to be positive from the beginning.\\\\

\noindent {\em Symmetries.} Since the equation of state is non-linear, the 
symmetries (\ref{symm1a})-(\ref{symm1b}) and (\ref{symm2a})-(\ref{symm2c}) 
do not apply to this analogous FLRW universe.

The third symmetry is enjoyed by an equation of the 
form~(\ref{eq:Friedmann}) with a {\em single} perfect fluid, or with a 
single fluid plus cosmological constant \cite{DussaultFaraoni}. In the 
present situation we necessarily have two fluids plus cosmological 
constant because eliminating one of the fluids necessarily implies setting 
$r=0$, which leads to losing the logistic equation.  As a conclusion, the 
third symmetry does not apply to Eq.~(\ref{logisticreduced}).

\subsection{Second analogy with comoving time}

As a second possible analogy, one can take instead a spatially flat 
($K=0$) universe with $\Lambda=0$ and energy density 
\be
\rho=\rho_0 \left(1-a\right)^2 \,; \label{density2}
\ee
imposing the covariant energy conservation equation yields 
\be
P=-\rho +\frac{2\rho_0}{3} a(1-a) 
\ee
and, eliminating $a$ with Eq.~(\ref{density2}), which yields $
a=1\pm \sqrt{ \rho / \rho_0 } $, 
one obtains the nonlinear equation of state
\be
P=-\frac{5}{3}\, \rho \mp \frac{2}{3} \sqrt{ \rho_0 \rho} \,.\label{eos2}
\ee
This time, the energy density is automatically non-negative and this is an 
acceptable analogy from the physical point of view. The FLRW universe 
analogous to the logistic equation 
is spatially flat, has zero cosmological constant, and is filled with the 
exotic fluid~(\ref{eos2}). Since $P<-\rho/3$, this universe is 
accelerated, as follows from the acceleration 
equation $\frac{\ddot{a}}{a} = -\frac{4\pi G}{3} \left( \rho +3P \right)$.\\\\
{\em Lagrangian and Hamiltonian.} The Lagrangian and Hamiltonian for this 
analogous universe
\begin{eqnarray}
L_2 &=&  3a\dot{a}^2 +8\pi G \rho_0 a^3 \left(1-a \right)^2 \,,\\
&&\nonumber\\
{\cal H}_2 &=&  3a\dot{a}^2 -8\pi G \rho_0 a^3 \left(1-a \right)^2 \,,
\end{eqnarray}
reproduce (apart from an irrelevant multiplicative constant) the 
Lagrangian~(\ref{L1}) and Hamiltonian~(\ref{H1}) for the 
logistic equation.\\\\ 

\noindent {\em Symmetries.} Since the equation of state is non-linear, the 
symmetries (\ref{symm1a})-(\ref{symm1b}) and (\ref{symm2a})-(\ref{symm2c}) 
do not apply to this analogous FLRW universe.  The remaining FLRW symmetry 
(\ref{symm3a})-(\ref{symm3c}) preserves the analogy 
(and, therefore, the logistic equation) provided that the equation of 
state~(\ref{eos2}) is maintained. This 	requirement implies that
\be
\pm \bar{\rho}^{3/2} + \sqrt{\bar{\rho}_0}\, \bar{\rho}=
\left( \pm \rho^{3/2} + \sqrt{\rho_0}\, \rho \right) \frac{ 
d\bar{\rho} }{d\rho} 
\ee
and, consequently,
\be
\int \frac{ d\bar{\rho} }{
\sqrt{\bar{\rho}_0}\, \bar{\rho} \pm \bar{\rho}^{3/2} } =  
\int \frac{d\rho}{
\sqrt{\rho_0}\, \rho \pm \rho^{3/2} } \,.
\ee
Computing the integral, one obtains  
\be
\frac{2}{ \sqrt{ \bar{\rho}_0} } \ln \left( \frac{ \sqrt{\bar{\rho}}}{
\sqrt{ \bar{\rho}} \pm \sqrt{ \bar{\rho}_0 }} \right) =  
\frac{2}{\sqrt{\rho_0} } \ln \left( \frac{ \sqrt{ \rho}}{
\sqrt{ \rho} \pm \sqrt{ \rho_0} } \right) \,,
\ee
which is conveniently  rewritten as
\be
\ln \left( 1 \pm \sqrt{ \frac{ \bar{\rho}_0}{ \bar{\rho}} }\right)  
= \sqrt{ \frac{\bar{\rho}_0}{\rho_0} } 
\ln \left( 1 \pm \sqrt{ \frac{\rho_0}{\rho} } \right) \,.
\ee
Using the fact that 
\be
\sqrt{ \frac{\bar{\rho}_0}{ \rho_0} }=\frac{ \bar{r} }{r} \,,
\ee
one obtains
\be
1 \pm \sqrt{ \frac{ \bar{\rho}_0}{ \bar{\rho}} }  
= \left( 1 \pm \sqrt{ \frac{\rho_0}{\rho} } \right)^{\bar{r}/r} 
\,,\label{eq:Andrea}
\ee
so that  
\be
1\pm \sqrt{ \frac{\rho_0}{\rho} } = \frac{ 1\pm 1 -f}{1-f} \,; 
\label{eq:Andrea2}
\ee
let us adopt the lower sign first. Then Eq.~(\ref{eq:Andrea}) yields
\be
-\frac{ \bar{f} }{1-\bar{f} } = \left( - \frac{ f}{1-f} \right)^{ 
\bar{r}/r} \,,\label{azzo2}
\ee
which, since the argument of the parenthesis in the right hand side is 
negative,  is well defined only if $\bar{r}/r \equiv n $ is an integer. If 
$n $ is even, Eq.~(\ref{azzo2}) has no solutions because the left hand 
side is negative while the right hand side is positive. If instead  
$n=2k+1$,  $k\in \mathbb{N}$, then~(\ref{azzo2}) reduces to
\be
f\rightarrow \bar{f} = \frac{ \left( \frac{f}{1-f} \right)^{2k+1} }{ 1+ 
\left( \frac{f}{1-f} \right)^{2k+1} } \,. \label{eq:45}
\ee
The sigmoid solution~(\ref{sigmoid}) of the logistic equation clearly 
enjoys this symmetry, which corresponds to the simple rescaling
\be
f =\frac{f_0 \, \mbox{e}^{\beta \, t}}{1+f_0 \, \mbox{e}^{\beta \, t} }    
\rightarrow \bar{f} =
\frac{\bar{f}_0 \, \mbox{e}^{\bar{ \beta} \, t}}{1+\bar{f}_0 \, 
\mbox{e}^{ \bar{\beta} \, t} }     
\ee
with
\begin{eqnarray}
\bar{f}_0 &=& f_0^{2k+1} \,,\\
&&\nonumber\\
\bar{\beta} &=& (2k+1) \beta \,,
\end{eqnarray}
which, alternatively, follows directly from the properties of the 
exponential function in the solution.

If we instead adopt the upper sign in Eq.~(\ref{eq:Andrea2}), we obtain
\begin{eqnarray}
\frac{ 2-\bar{f} }{1-\bar{f} } &=& 
\left( \frac{ 2-f}{1-f}\right)^{ \bar{r}/r} \equiv \gamma \,,\\
&&\nonumber\\
\bar{f} &=& \frac{2-\gamma}{1-\gamma} \,,
\end{eqnarray}
from which it follows that the solution~(\ref{sigmoid}) is mapped into
\be
\bar{f}= 1+ \frac{1}{1- \left( 2 + f_0 \mbox{e}^{\beta \, t} 
\right)^{\bar{r}/r} }\,. \label{eq:51}
\ee
Certain symmetries of the logistic equation are hidden and are made 
manifest by the analogy with cosmology. Otherwise, it is not obvious how 
to look for such symmetries, either in the equation or in its solutions. 
This is the case, for example, of the symmetries described by 
Eq.~(\ref{eq:45}) or Eq.~(\ref{eq:51}), which apparently are not found in 
systematic searches for symmetries.

\subsection{First analogy with conformal time}

One can rewrite the Friedmann equation~(\ref{eq:Friedmann}) using 
conformal, 
instead of comoving, time. The 
conformal time $\eta$ is defined by $dt \equiv ad\eta$ \cite{Wald}. Assume 
that 
there is a 
single perfect fluid with constant equation of state $P=w\rho$ and 
$\Lambda=0$. Then the  use of 
conformal time turns the combined Friedmann 
equation~(\ref{eq:Friedmann}) and acceleration equation 
$\frac{\ddot{a}}{a} = 
-\frac{4\pi G}{3} \left(3w+1 \right)\rho $   into 
\be
\frac{a_{\eta\eta}}{a}+ \left( c-1\right) \left( \frac{a_{\eta}}{a} 
\right)^2 +cK=0 \,,\label{intermediate}
\ee
where $c= (3w+1)/2$. Using the new variable $u\equiv a_{\eta}/a$, 
Eq.~(\ref{intermediate}) is reduced to the Riccati equation \cite{myAmJP}
\be
u_{\eta}+cu^2 +cK=0 \,.\label{Riccati}
\ee
This reduction is a known way of solving the Einstein-Friedmann equation 
for a universe filled with a perfect fluid with constant equation of 
state, which is an alternative to the more standard solution method using 
quadratures \cite{Barrow,myAmJP}. It has been used as the basis for 
several applications in cosmology 
\cite{Rosu1,Rosu2,Rosu3,Rosu4,Rosu5,Rosu6,Rosu7,Harko,Schuch,LRRiccati}.

The logistic equation can also be reduced to a Riccati equation 
\cite{Schuch}. From a broader point of view, this is not the only 
integrable first order equation that connects with FLRW universes, 
especially those filled with scalar fields. For example, the solution of 
the Riccati equation~(\ref{Riccati}) is a hyperbolic tangent, while 
cosmological models based on a phantom scalar field expressed by an 
hyperbolic tangent are known \cite{ArefevaKoshelevVernov06}. Another first 
order equation used to integrate a cosmological model based on a scalar 
field with exponential potential appears in \cite{SalopekBond90}. Another 
connection between scalar field cosmology and the Abel equation of the 
first kind was investigated in \cite{YurovYurov10}. More generally, 
connections between the Einstein-Friedmann equations and the 
Ermakov-Pinney equation \cite{Hawkins:2001zx} and the Schr\"odinger 
equation are known \cite{Fomin:2018uql,BarbosaCendejas:2010td}.

Proceeding with the Riccati equation~(\ref{Riccati}) and setting 
\be
f(t) \equiv \frac{1}{2} \left[ g(t)+1 \right] \,,
\ee
Eq.~(\ref{logistic}) becomes 
\be
\dot{g}+\frac{r}{2} \, g^2 -\frac{r}{2} =0 \,,\label{oftheform}
\ee
which is of the form~(\ref{Riccati}) with $c=r/2$, $K=-1$. We have then 
the formal analogy $\left( \eta, u(\eta) \right) \rightarrow \left(t, g(t) 
\right)=\left( t, 2f(t)-1 \right) $ provided that  $K=-1$ and 
$c=r/2=(3w+1)/2$, which gives 
$w=(r-1)/3$. Two special cases correspond to classic solutions of FLRW 
cosmology \cite{Wald, Carroll, Liddle,KolbTurner}:  $r=1$ corresponds to a 
dust 
(zero pressure fluid), while 
$r=2$ corresponds to a radiation fluid, both of which are solved exactly  
\cite{Wald}.

The analogy implies that 
\be
\frac{a_{\eta}}{a}=u = g(\eta)=2f(\eta)-1 \,;
\ee
using the solution~(\ref{sigmoid}) and integrating gives the scale factor 
of the analogous universe in conformal time
\be
a(\eta) = a_0 \, \mbox{e}^{-\eta} \left( 1+f_0 \mbox{e}^{r\eta} 
\right)^{2/r} \,. \label{azzo3}
\ee
The comoving time $t$ is obtained from the conformal time by integration, 
producing the hypergeometric function 
\be
t=\int ad\eta= -a_0 \, \mbox{e}^{-\eta} \, _2F_1 \left( -\frac{2}{r}, 
-\frac{1}{r}, \frac{r-1}{r}, -f_0 \, \mbox{e}^{r\eta} \right) +t_0 
\,,\label{hypergeometric}
\ee
where $t_0$ is an integration constant. Equations~(\ref{azzo3}) and 
(\ref{hypergeometric}) constitute a parametric representation of the 
solution $a(t)$ in terms of comoving time with $\eta$ as  
the parameter. Unfortunately, this representation is too cumbersome for 
practical uses in most situations and it is more convenient to use 
$a(\eta)$ instead.\\\\

\noindent {\em Lagrangian and Hamiltonian.} A Lagrangian for the 
Riccati equation~(\ref{oftheform}) is\footnote{The Lagrangian formalism 
for second order Riccati equations different from the present one was 
studied in Ref.~\cite{LRRiccati}.} 
\be
L_R \left( g, \dot{g} \right) =  \dot{g}^2 +\frac{r^2}{4} \left( 
g^2-1\right)^2 \,,
\ee
which does not depend explicitly on $\eta$, hence the corresponding 
Hamiltonian 
\be
{\cal H}_R = \dot{g}^2- \frac{r^2}{4} \left( g^2-1\right)^2 
\ee
is conserved. Choosing zero value for this constant and the negative sign 
of the root of ${\cal H}=0 $ reproduces Eq.~(\ref{oftheform}).\\\\

\noindent {\em Symmetries.} {\em A priori} the 
symmetries~(\ref{symm1a})-(\ref{symm1b}) 
and (\ref{symm2a})-(\ref{symm2c}) should not apply to the equation 
$u_{\eta}+ cu^2-c=0$ because this implies $K=-1$, However, the first 
symmetry  translates into 
\be
u \rightarrow  \bar{u}=\frac{1}{u} \,,
\ee
which leaves this equation invariant. As expected, the second FLRW 
symmetry  does not 
translate into useful symmetries of the Riccati equation. The 
third symmetry, valid for spatially flat FLRW universes,  does not apply 
 here because $K$  is fixed to the value $-1$.

\subsection{Second analogy with conformal time}

Instead of the analogy $\left( \eta, u(\eta) \right)\rightarrow \left( t, 
f(t) \right)$, let us consider now the more direct analogy 
$\left( \eta, a(\eta) \right)\rightarrow \left( t, f(t) \right)$. The 
Friedmann equation in conformal time reads
\be
\left( \frac{a_{\eta} }{a} \right)^2 = \frac{\Lambda a^2}{3} +\frac{8\pi G 
}{3} \, \rho a^2 -K \,, \label{inconfotime2}
\ee
which is analogous to Eq.~(\ref{logisticreduced}) if we make the 
correspondence
\begin{eqnarray}
K &=& - r^2 \,, \label{urca1}\\
&&\nonumber\\
\Lambda &=& 3r^2\,,\label{urca2}\\
&&\nonumber\\
\rho &=& -\frac{3r^2}{4\pi G} \, \frac{1}{a} \equiv -\frac{\rho_0}{a} 
\,.\label{azzo}
\end{eqnarray}
The energy density~(\ref{azzo}) is negative, which would lead to rejecting  
this analogy on purely physical grounds.\\\\
{\em Lagrangian and Hamiltonian.} The usual Lagrangian can be used, 
provided that the cosmological constant is treated as an extra perfect 
fluid with energy density $\rho_{\Lambda}=\Lambda/(8\pi G)$ added to the 
usual perfect fluid, which yields
\be
L_3= \frac{a_{\eta}^2}{a} +\frac{8\pi G}{3} \, \rho a^3 -Ka +\frac{\Lambda 
a^2}{3} \,,
\ee
and the Hamiltonian
\be
{\cal H}_3 = \frac{a_{\eta}^2}{a} -\frac{8\pi G}{3} \, \rho a^3 +Ka 
-\frac{\Lambda a^2}{3} \,.
\ee
Using the identifications~(\ref{urca1})-(\ref{azzo}), the logistic  
equation is equivalent to ${\cal H}_3=0$ (with the choice of the positive 
sign when taking the square root of both sides).\\\\

\noindent {\em Symmetries.} Since the spatial sections of this universe 
are spatially curved, the symmetries (\ref{symm1a})-(\ref{symm1b}) and 
(\ref{symm2a})-(\ref{symm2c}) do not apply to this analogous FLRW 
universe. The third symmetry is enjoyed by an equation of the 
form~(\ref{eq:Friedmann}) with a {\em single} perfect fluid, or with a 
single fluid plus cosmological constant \cite{DussaultFaraoni}. In our 
case, eliminating the second fluid (or $K$-fluid) by setting $K=0$ means 
setting $r=0$ and losing the logistic equation altogether. Therefore, the 
third symmetry does not apply to Eq.~(\ref{inconfotime2}).

\section{Discussion and conclusions}
\setcounter{equation}{0}
\label{sec:4}

On the basis of the previous analogies, we can extend the knowledge of 
solutions of the Einstein-Friedmann equations to those of the logistic 
equation. In particular, certain recent results on the analytic solutions 
of the Einstein-Friedmann  
equations \cite{Chen0,  Chenetal2015a, Chenetal2015b} can be immediately 
transposed to the logistic equation.  First, Ref.~\cite{Chenetal2015b} 
contains an explicit proof 
that all solutions of the Friedmann equation~(\ref{eq:Friedmann}) are 
roulettes. Therefore, one concludes without effort that all solutions of 
the logistic equation are also roulettes. [A roulette is the trajectory in 
two dimensions  described by a point that lies on a curve rolling without 
slipping on another given curve.] 

Second, one wonders under which conditions it is possible to 
obtain analytic solutions of the Einstein-Friedmann equations in terms 
of elementary functions. This non-trivial question is answered 
in Ref.~\cite{Chen0} using Chebysev's theorem of integration  
\cite{Chebysev,MarchisottoZakeri}. The transposition of this result to the 
logistic equation does not provide new information, since the solution of 
the latter is the well known sigmoid.

Third, one can deduce effective Lagrangians and Hamiltonians for the 
logistic equation (and also for a special Riccati equation appearing in 
the cosmological analogy using conformal time). The most important 
consequence of the Lagrangian formulation is probably that these systems 
admit a conserved energy function. Parallel to the fact that the 
dynamics of FLRW (and, in a wider context, of general relativity) is 
constrained, 
this conserved energy is forced to have zero value. Furthermore, these 
Lagrangian and Hamiltonian are obtained by squaring a first order 
differential equation, thus introducing one extra mode with respect to the 
original equation. Hence, a sign choice must be made in order to reproduce 
the correct equation that we started from (see \cite{focus} for a more 
comprehensive discussion of this procedure). Certain symmetries of the 
Einstein-Friedmann equations unveil hidden symmetries of the logistic 
equation. It is currently not clear whether these symmetries are useful 
for applications of the logistic equation.

It seems that the cosmology side of the analogy has less to gain. 
However, viewing the same problem from a different angle is often 
interesting {\em a priori}. For example, it is interesting to see the 
evolution of the scale factor $a(t)$ for the phantom fluid examined here 
on par with the evolution of a population described by Verhulst with the 
logistic equation.

\begin{acknowledgments} 

We thank two referees for constructive comments. This work is supported, 
in part, by the Natural Sciences \& Engineering Research Council of Canada 
(Grant No. 2016-03803 to V. Faraoni) and by Bishop's University. A. Giusti 
is supported by the European Union's Horizon 2020 research and innovation 
programme under the Marie Sk\l{}odowska-Curie Actions (grant agreement No. 
895648 -- CosmoDEC). His work has been carried out in the framework of the 
activities of the Italian National Group for Mathematical Physics [Gruppo 
Nazionale per la Fisica Matematica (GNFM), Istituto Nazionale di Alta 
Matematica (INdAM)].

\end{acknowledgments}

% Use \appendix* if there is only one appendix.
%\appendix
%\section{Derivation of Eq.~(\ref{wdot}) }
%\renewcommand{\theequation}{A.\arabic{equation}}
%\setcounter{equation}{0}

%\bibliography{simplified}

\end{document}